\documentclass[12pt]{spieman}
\usepackage{graphicx}
\usepackage{amsmath,amsfonts,amssymb}
\usepackage{fixltx2e}
\usepackage{pdfpages}
\usepackage{multirow}
\usepackage{setspace}
\usepackage{tocloft}
\newcommand\adeg{\mbox{$^\circ$}}%
\makeatletter
\newcommand*{\rom}[1]{\expandafter\@slowromancap\romannumeral #1@}
\makeatother
\def\ion#1#2{#1$\;${\small\rm\rom{#2}}\relax}
\newcommand\amin{\mbox{$^\prime$}}%
\newcommand\asec{\mbox{$^{\prime\prime}$}}%
\newcommand\lya{\mbox{Ly$\alpha$}}%

\title{The Colorado Ultraviolet Transit Experiment (CUTE): A dedicated cubesat mission to study exoplanetary mass loss and magnetic fields} 

\author[a,*]{Brian T. Fleming}
\author[a]{Kevin France} 
\author[a]{Nicholas Nell} 
\author[a]{Richard Kohnert}
\author[a]{Kelsey Pool} 
\author[a]{Arika Egan} 
\author[b]{Luca Fossati}
\author[c]{Tommi Koskinen}
\author[d]{Aline A. Vidotto} 
\author[e]{Keri Hoadley} 
\author[a]{Jean-Michel Desert} 
\author[f]{Matthew Beasley}
\author[g]{Pascal Petit}
\affil[a]{University of Colorado, Laboratory for Atmospheric and Space
  Physics, UCB 600, Boulder, CO, 80309, USA}
\affil[b]{Austrian Academy of Sciences, Space Research Institute, Graz, Austria}
\affil[c]{University of Arizona, Lunar and Planetary Laboratory,
  Tuscon, AZ, 85721, USA}
\affil[d]{Trinity College Dublin, School of Physics, University of
  Dublin, Dublin, Ireland}
\affil[e]{California Institute of Technology, Pasadena, CA, 91125, USA}
\affil[f]{Planetary Resources LLC, Redmond, WA, 98052, USA}
\affil[g]{Universit\'{e} de Toulouse, IRAP, CNRS, CNES, UPS, Toulouse, France}

\cftpagenumbersoff{figure}
\cftpagenumbersoff{table} 

\begin{document} 

\maketitle

\begin{abstract}
The Colorado Ultraviolet Transit Experiment (CUTE) is a near-UV (2550
-– 3300 \AA ) 6U cubesat mission designed to monitor transiting hot
Jupiters to quantify their atmospheric mass loss and magnetic
fields. CUTE will probe both atomic (Mg and Fe) and molecular (OH)
lines for evidence of enhanced transit absorption, and to search for evidence of early ingress due to bow shocks ahead of the planet's orbital motion. As a dedicated mission, CUTE
will observe $\gtrsim$ 100 spectroscopic transits of hot Jupiters over a nominal seven month mission. This
represents the equivalent of $>$ 700 orbits of the only other
instrument capable of these measurements, the {\em Hubble Space
Telescope}. CUTE efficiently utilizes the available cubesat volume by means of an
innovative optical design to achieve a projected effective area of
$\sim$ 28
cm$^{2}$, low instrumental background, and a spectral resolving power of R
$\sim$ 3000 over the primary science bandpass. These performance
characteristics enable CUTE to discern transit depths between 0.1 -- 1$\%$
in individual spectral absorption lines. We present the CUTE optical and mechanical design, a summary of the science motivation and expected results, and an overview of the projected fabrication, calibration and launch timeline. 
\end{abstract}


{\noindent \footnotesize\textbf{*}Brian T. Fleming,  \linkable{Brian.Fleming@lasp.colorado.edu} }

\begin{spacing}{1}   

\section{Introduction}

Small satellite missions enable the study of transient phenomena over extended time periods in a manner not feasible
for large, multi-purpose space observatories such as $HST$. Long-term
monitoring of exoplanets in short period orbits (1 -- 5 days) provides
a unique opportunity to observe the interplay between planetary
atmospheres and the host star.  Repeatable, near-ultraviolet (NUV; 2550 -- 3300 \AA ) transit spectroscopy enables quantification of the
atmospheric mass-loss and potentially planetary magnetic fields. The Colorado
Ultraviolet Transit Experiment (CUTE) is the first NASA funded
UV/O/IR astrophysics cubesat, and second overall (the first being the
University of Iowa x-ray mission HaloSat). CUTE leverages a compact optical design
in a 6U cubesat form factor to provide a high
efficiency NUV spectrograph dedicated to
monitoring the spectral properties of hot Jupiter atmospheres during
transit. 

The typical transit depth of hot Jupiters at visible wavelengths
is $\sim$ 1$\%$, however the atmospheres of short-period planets
may be inflated to several planetary radii, resulting in transit depths
of 3 -- 10$\%$ in specific spectral
tracers \cite{Fossati10, Haswell12}. Some systems present late
egresses indicative of the presence of a cometary tail due to
the planetary atmosphere being dragged by stellar wind and radiation as it extends beyond the Roche lobe. It is also possible that these planets will exhibit an early
ingress indicative of atmospheric material preceding the planet in
its orbit. NUV transit
measurements of an early ingress in the close-orbiting WASP-12 system were interpreted as due to an
optically thick bow shock supported by either a planetary magnetic
field \cite{Vidotto11a,Nichols15} or a high mass-loss rate \cite{Turner16}.

Observing time resources on orbiting observatories are insufficient to monitor enough transiting
systems to statistically characterize the interplay between short-period, massive
planets and their host stars. Broadband visible/NIR light curves, such as those
generated by TESS or Kepler, are not sensitive to atmospheric
tracers. The observation time necessary to establish a complete transit
lightcurve for multiple systems, with multiple visits to each system to
check for variability, is too costly for a shared flagship
resource such as $HST$. The time to completely map a
transit for a single system is determined not by the sensitivity of
the observatory, but rather by the length of time the planet is in-transit and the observational time available (set
by the orbit of the spacecraft). As a dedicated satellite, CUTE will
carry out the first survey of NUV spectral lightcurves of short-period
exoplanets. 

\begin{figure}
\begin{center}
   \begin{tabular}{c}
\includegraphics[angle=0,width=0.65\textwidth,trim={0.1in 0.1in 0.1in
  0.1in},clip]{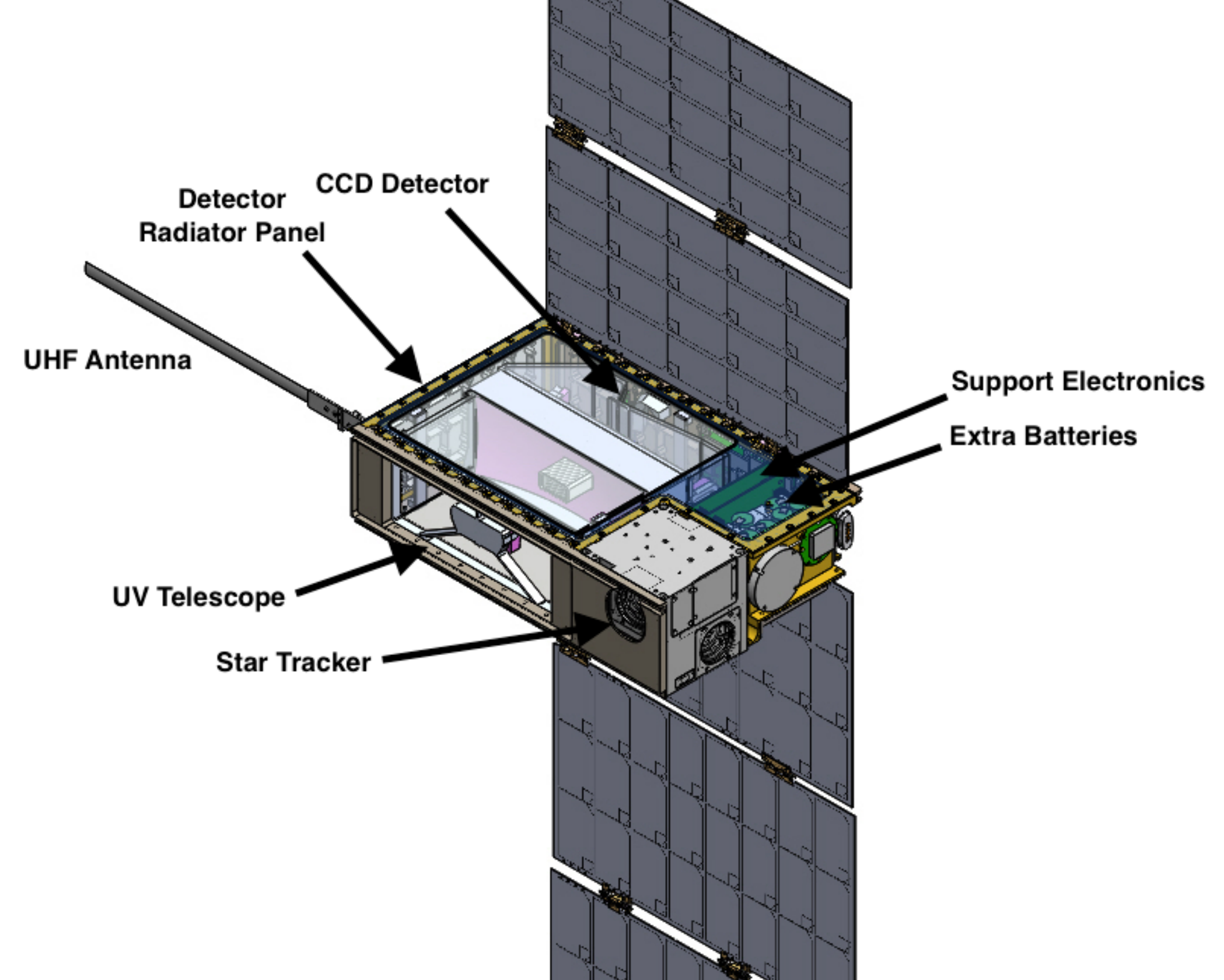}
 \end{tabular}
   \end{center}
\caption[Trackfits] 
 { \label{fig-render} 
Transparent rendering of the CUTE spacecraft.}
\end{figure}

The spectral resolving power of CUTE is comparable to the G230L mode of $HST$-COS (R $\approx$ 3000) over a
bandpass that covers critical atomic and molecular tracers such as \ion{Mg}{1},
\ion{Mg}{2}, \ion{Fe}{2}, and OH that are inaccessible from the
ground. A compact design that maximizes throughput makes CUTE
sensitive to transit depths of $>$ 1.2$\%$ in \ion{Mg}{2} to greater
than 3$\sigma$ confidence in a single
transit for the median planet in the preliminary CUTE target list, and
$>$ 0.7$\%$ in the continuum. Folded over multiple transits, the continuum
sensitivity reaches down to $<$ 1$\%$ transit depths for all targets, and $<$
0.1$\%$ for HD 209458b. In this
paper, we will outline the motivation for CUTE (\S\ref{sciback}), provide an overview of the
CUTE instrument design (\S\ref{Instrument}), performance
(\S\ref{performance}), observation
strategy (\S\ref{sci_op}), and the integration and testing (I $\&$ T) timeline
to launch (\S\ref{schedule}). 
  
\section{Scientific Background}\label{sciback}
\begin{figure}
   \begin{center}
   \begin{tabular}{c}
\includegraphics[angle=0,width=0.85\textwidth,trim={0.1in 0.1in 0.1in
  0.1in},clip]{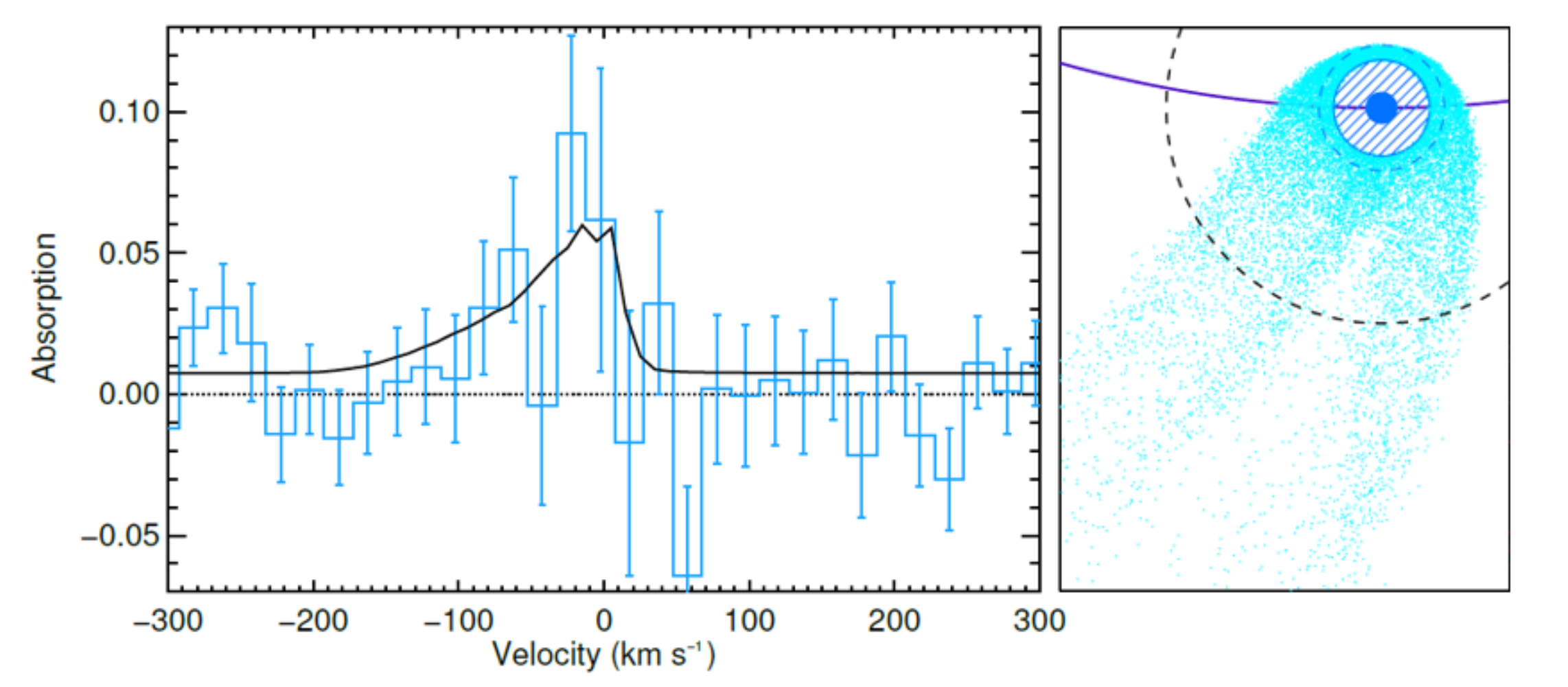}
  \end{tabular}
   \end{center}
\caption[Trackfits] 
 { \label{fig-ingress} 
(Left) HST-STIS observations of neutral magnesium mass-loss from HD
209458b \cite{Vidal13} and (right) the hydrodynamic mass-loss model from Bourrier et al. (2014) \cite{Bourrier14}.}
\end{figure}

Exoplanets in short-period orbits provide a laboratory to
study extreme mass-loss from planetary systems with a small, dedicated
satellite. To date, there are only a handful ($<$ 10) of published UV measurements
of hot Jupiter atmospheres carried out with $HST$. The detection of hydrodynamic escape using UV
measurements of \lya, \ion{O}{1}, \ion{C}{2}, \ion{Si}{3}, and
\ion{Mg}{1} \cite{Vidal04, Vidal13, Linsky10} led to the
development of multiple 1D and 3D models of the upper atmospheres of
short-period planets (Fig.~\ref{fig-ingress}) \cite{Koskinen07, Koskinen13a, Koskinen13b, Bourrier13}. The
interpretation of many far-UV (FUV) transit measurements is controversial,
however, due in part to the small-number statistics and
uncertain spatial and temporal uniformity of the stellar FUV chromospheric emission \cite{Fossati15}.

 The NUV provides both
higher stellar flux and a better understood spectral intensity
distribution relative to the FUV\cite{Haswell12}, as well as a diverse set of neutral, ionized, and molecular spectral
lines that trace heavy elements \cite{Fossati10}. These elements would normally condense into
clouds in the lower atmosphere, however they may be dragged to the upper
atmosphere (and thus escape) by a sufficient mass flow rate of
hydrogen \cite{Koskinen13b}. With hydrodynamic models at hand, the CUTE
data will provide a measure of the global mass loss rate, as well as
the elemental abundances of species like Mg and Fe. These NUV
spectral lines probe the interaction between the planetary exosphere
and the stellar wind, enabling CUTE to also investigate the
interplanetary media of the host stars. Water is probably the most important infrared active species in hot Jupiter atmospheres where it regulates the temperatures in the middle and upper atmospheres.
OH is the primary dissociation product of water; detection of OH absorption in an exoplanet atmosphere 
would provide critical constraints on the fate of water in exoplanet atmospheres.  
OH can also act as a coolant, and detecting it with substantial abundances would provide new clues to the energy balance of the upper atmosphere.

Models of exoplanet magnetic fields predict radio emission \cite{Vidotto12} and FUV auroral emission \cite{Yelle04} may be detectable from Jovian-mass planets in the solar neighborhood, however, these signals have not been conclusively detected \cite{Bastian00,France10,Lecavelier13}, 
leaving the details of exoplanetary magnetism essentially unconstrained. Previous NUV transit observations of WASP-12b indicate an early
ingress that has been interpreted in a number of theoretical
scenarios, including an star-leading accretion
stream supported by hydrodynamic mass loss \cite{Turner16}, a plasma
torus created by satellites of WASP-12b \cite{Kislyakova16}, or a
magnetically supported bow shock upstream of the planet's orbit
\cite{Vidotto10,Llama11}. Subsequent $HST$ transit measurements of
WASP-12b, however, returned ambiguous results on the early ingress,
suggesting potential stellar or planetary variability \cite{Haswell12,Nichols15}. CUTE will provide the sensitivity and
temporal resolution necessary to discern between these scenarios if the
early ingress is detected in a significant number of systems. Observing
between 6 and 10 transits per system will enable CUTE to track
variability and identify potential systematic errors. When combined with ground-based spectropolarimetric measurements of the
stellar magnetic field, CUTE could potentially provide a
detection and measurement of the planetary magnetic field itself.

The CUTE satellite will observe multiple transits of multiple hot
Jupiter systems over the seven month
nominal mission timeline. Ground-based spectropolarimetric
observations will be carried out in the northern and southern
hemispheres mainly with ESPaDOnS, Neo-NARVAL\cite{NARVAL}, and
HARPSpol, which members of the CUTE science team have institutional access to. We are currently targeting launching CUTE
into a sun-synchronous orbit (SSO) to observe at least 12
systems with ten observed transits each. The mission could be
extended until de-orbit depending on the health of the
instrument at the end of the nominal mission. 

\section{CUTE Instrument Overview}\label{Instrument}

A velocity resolution of $\lesssim$ 150 km s$^{-1}$ and an effective area
sufficient to obtain 3$\sigma$ transit detections to the geometric
planet size transit depth ($\sim$ 1$\%$) for all targets is
desired. The designated spectral resolution enables CUTE to
potentially resolve individual atmospheric absorption lines with
widths on the order of those observed in HD 209458 \cite{Bourrier14}
on high signal-to-noise targets (e.g. KELT-9b). Owing to the overlap
of many neutral and low-ionization atomic lines in the CUTE bandpass,
the primary mass-loss and lightcurve morphology analysis can be
carried out at reduced resolution in the event of unanticipated
optical performance degradation. The CUTE system is designed to exceed these performance specifications by
maximizing the telescope collecting area and utilizing high quality
optical elements, including UV-grade mirrors and a focusing, ultra-low scatter
holographically ruled grating. The aberration correcting grating and
fold mirror result in a spectrum with low cross-dispersion flaring, minimizing the
spectral extraction region on the detector (and thus the background
equivalent flux), while optimizing the spectral resolving power to
nearly the maximum possible (as limited by the detector pixel size) for the detector/bandpass combination selected. The rectangular aperture of CUTE
is a new innovation for a cubesat and represents more than three times
the collecting area possible with a circular telescope. All other CUTE systems have flight heritage based on similar components flown on previous or upcoming missions. We present in the following sections an overview of the optical and mechanical design of
the telescope and spectrograph (\S\ref{telescope}), a description of the detector focal plane
array (\S\ref{detector}), and a breakdown of the overall
instrument performance (\S\ref{performance}). 

\subsection{Optical Design}\label{telescope}

\begin{figure}
   \begin{center}
   \begin{tabular}{c}
 \includegraphics[width=0.5\textwidth,angle=0,trim={0in 0.0in 0.0in 0.0in},clip]{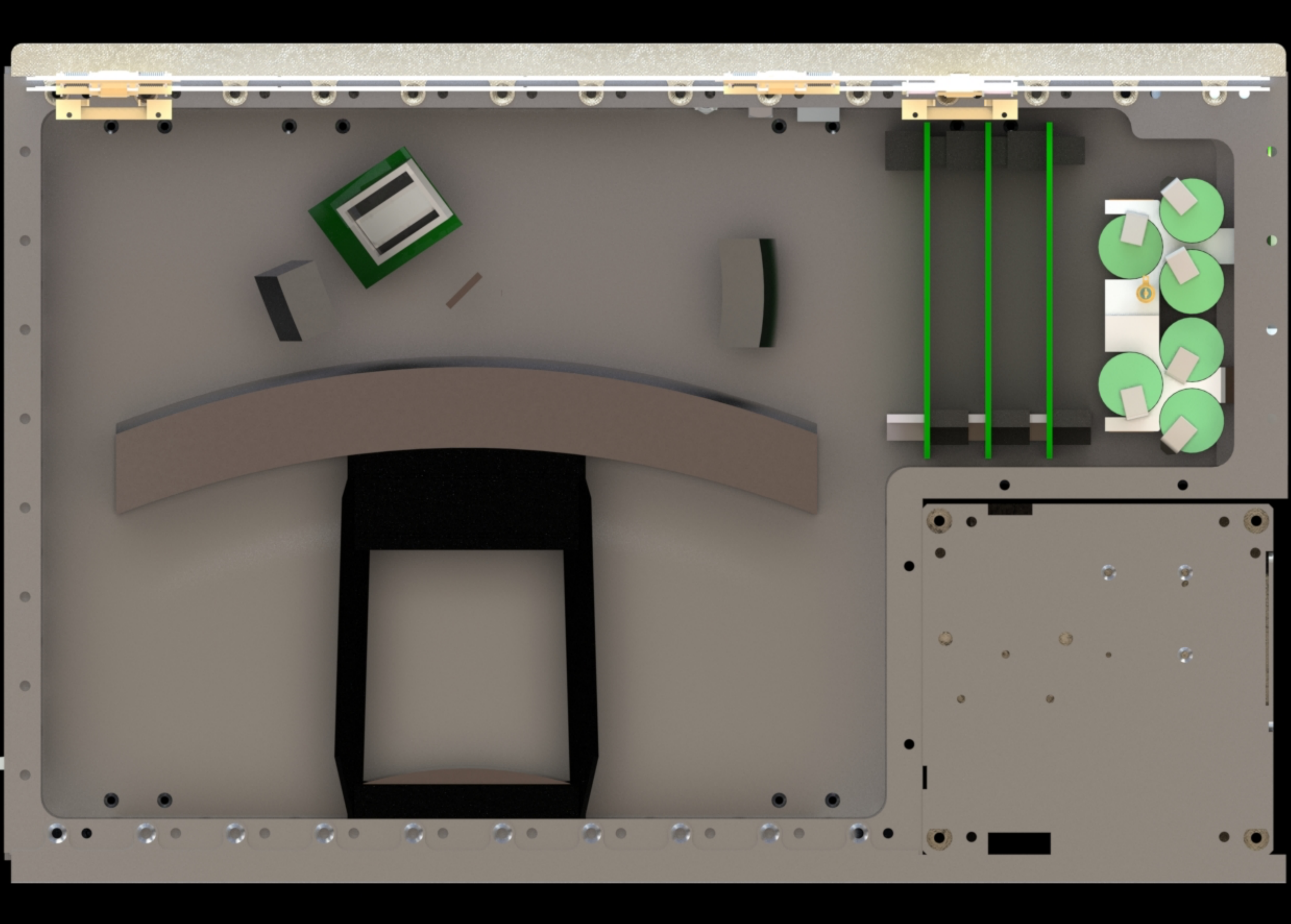} 
   \end{tabular}
   \end{center}
   \caption[example] 
  { \label{fig-render} 
CAD rendering of the CUTE optics, with
a conceptual cartoon of the invar tower and telescope baffle,
inside the BCT 6U spacecraft. }
   \end{figure}

\begin{table}[ht]\scriptsize
\caption{CUTE Parameters}
\label{tbl-optics}\vspace{-0.00in}
\begin{center}
\begin{tabular}{ll}
\hline \hline
\multicolumn{2}{c}{{\bf CUTE Instrument Summary}}\\
\hline
Focal Ratio & F/5.5 \\
Resolving Power (R) at 3000 \AA\ & 3700 \\
 Waveband (Total) & 2515 - 3335 \AA\ \\
 Field of View & 23.0\amin\ \\
Instrument Platescale & 208\asec\ mm$^{-1}$ \\
Instrument Effective Area (3000 \AA) & 29.3 cm$^{2}$ \\
Total Focusing Length & 482.7 mm \\
Total Instrument Length & 195.0 mm \\
\hline
\multicolumn{2}{c}{{\bf Cassegrain Parameters}}\\
\hline
Primary Dimensions & 200 $\times$ 80 mm \\
 Secondary Dimensions & 68 $\times$ 26 mm \\
Primary Radius & 300 mm \\
 Secondary Radius & -129.6 mm \\
\hline
\multicolumn{2}{c}{{\bf Spectrograph Parameters}}\\
\hline
Blaze Wavelength & 2800.0 \AA\ \\
Incident Angle ($\alpha$) & 8.7\adeg\ \\
Output Angle ($\beta$) & 20.6\adeg\ \\
Line Density & 1714 gr mm$^{-1}$ \\
Grating Radius & 86.1 mm  \\
Grating Dimensions & 31 $\times$ 31 mm  \\
Fold Mirror Dimensions & 25 $\times$ 25 mm  \\
Detector Dimensions & 27.6 $\times$ 6.9 mm  \\
\hline
\end{tabular}\\
\end{center}
\end{table}

The CUTE aperture is rectangular in shape, featuring a 20 $\times$
8 cm, F/0.75 parabolic primary mirror feeding an F/2.6 classical Cassegrain
telescope. We reference the F-number for this rectangular aperture to
the longer axis of the telescope, which is the cross-dispersion
dimension of the spectrograph. The large telescope aperture is
made possible by the Blue Canyon Technologies
(BCT) XB1 spacecraft bus, which provides critical systems such as
power, command and data handling, communications, and attitude control
in a compact package that requires less than 2U of the 6U spacecraft. The hyperbolic secondary mirror will be cantilevered off
of an invar central spire attached to the primary mirror baseplate via
the primary aperture (Fig.~\ref{fig-render}). The telescope and mount structure will be
fabricated, aligned and vibration tested by Nu-Tek Precision Optics,
with redundant vibration testing done by CU after delivery to verify
alignment stability under launch loads.

The beam is folded 90\adeg\ by a 5 $\times$ 2.5 mm flat
mirror positioned 10 mm before a 200 $\mu$m $\times$ 3.5 mm (80\asec\
$\times$ 1400\asec ) slit at the Cassegrain focus. The slit
assembly will be polished and angled 45\adeg\ about the slit axis
(reducing the projected slit width to $\sim$ 100 $\mu$m) to
reflect the remainder of the field onto a ground service aspect
camera for alignment to the BCT spacecraft during integration. The beam is
diffracted, magnified and focused by a spherical R=86.1 mm, 1714 gr/mm aberration correcting
ion-etched holographic grating of the type used in $HST$-COS\cite{Green12} and
ruled by Horiba J-Y. A
second fold mirror with an R=300
mm cylindrical shape about the
cross-dispersion axis adds an additional layer
of aberration correction and positions the focal
plane to maximize the volume available for the detector. There are
some polarization effects induced by the grating and fold mirror that
are not anticipated to influence the science, but will still be
measured during testing and are accounted for in throughput
calculations (\S\ref{aeff}). The final
beam focal ratio is F/5.5, yielding a detector platescale of 208\asec\
mm$^{-1}$. 

\begin{figure}
   \begin{center}
   \begin{tabular}{c}
\includegraphics[height=0.49\textwidth,angle=0]{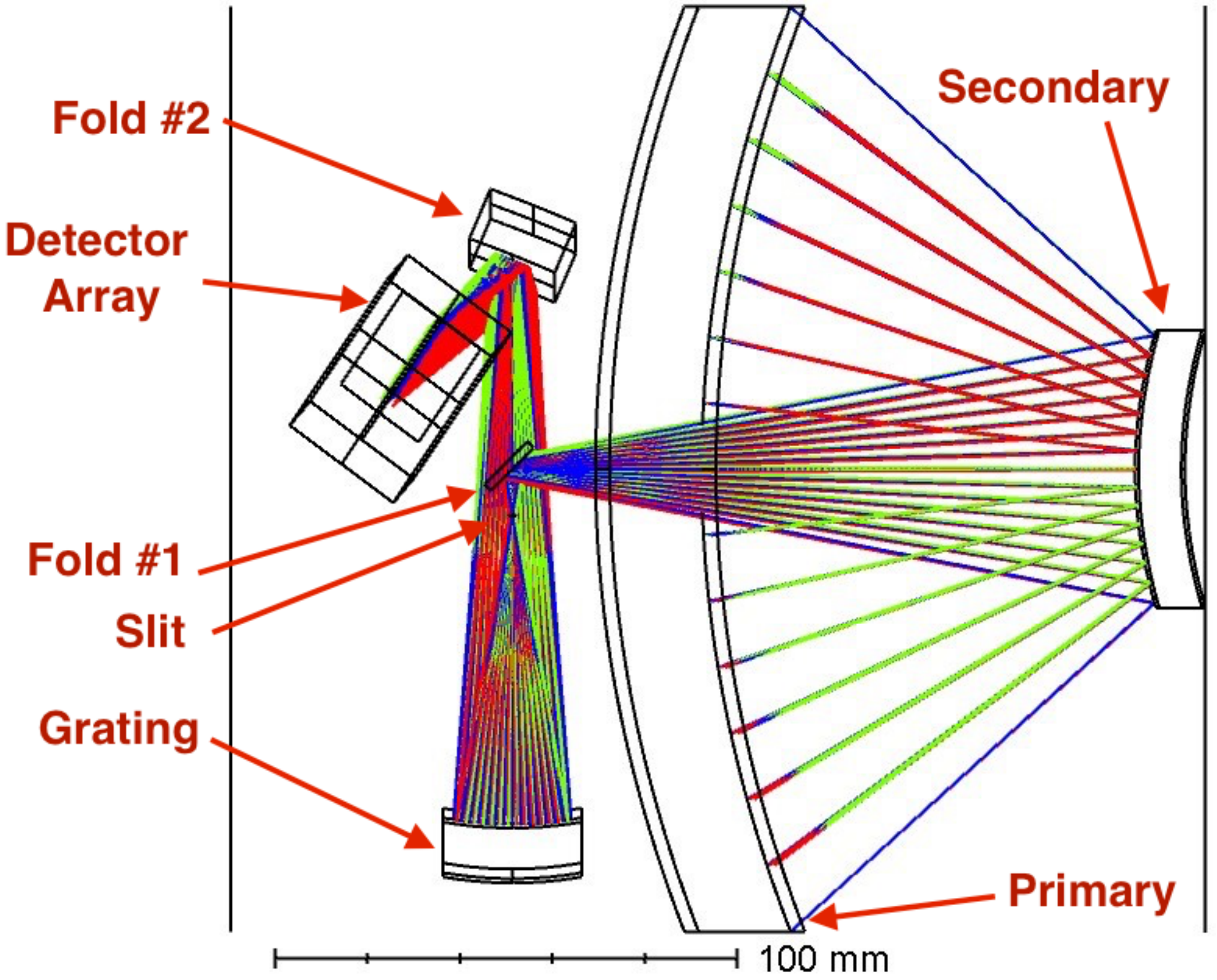}
   \end{tabular}
   \end{center}
   \caption[example] 
  { \label{fig-trace} 
Annotated raytrace of the CUTE spectrograph. The vertical black lines denote the
front and back walls of the BCT spacecraft. The dispersion
direction is out of the page. The detector footprint is the inner
surface, with an outer surface showing the footprint with mount.}
   \end{figure} 

Nu-Tek will fabricate a
mounting and metering assembly that will attach to the BCT spacecraft
via the primary mirror baseplate and feature mounts for the spectrograph optics and slit assembly,
while the grating and fold mirror will be installed and aligned at
CU. Blackened baffles will be fabricated into the optical structure to suppress
stray light between all optics (\S\ref{schedule}). There has been
recent progress in the fabrication of vacuum-safe baffles using 3D
printing, enabling more complex designs that will increase the stray
light rejection \cite{Mccandlissspie17}. We will investigate the feasibility of using these
materials and methods for the CUTE baffle system. A
raytrace of the optical system is presented in Figure~\ref{fig-trace}
and the optical prescription is presented in
Table~\ref{tbl-optics}.

\subsection{Detector System}\label{detector}
\begin{figure}
\begin{center}
\begin{tabular}{c}
\includegraphics[height=0.3\textwidth,angle=0,trim={0.05in 0.15in 0.3in 0.1in},clip]{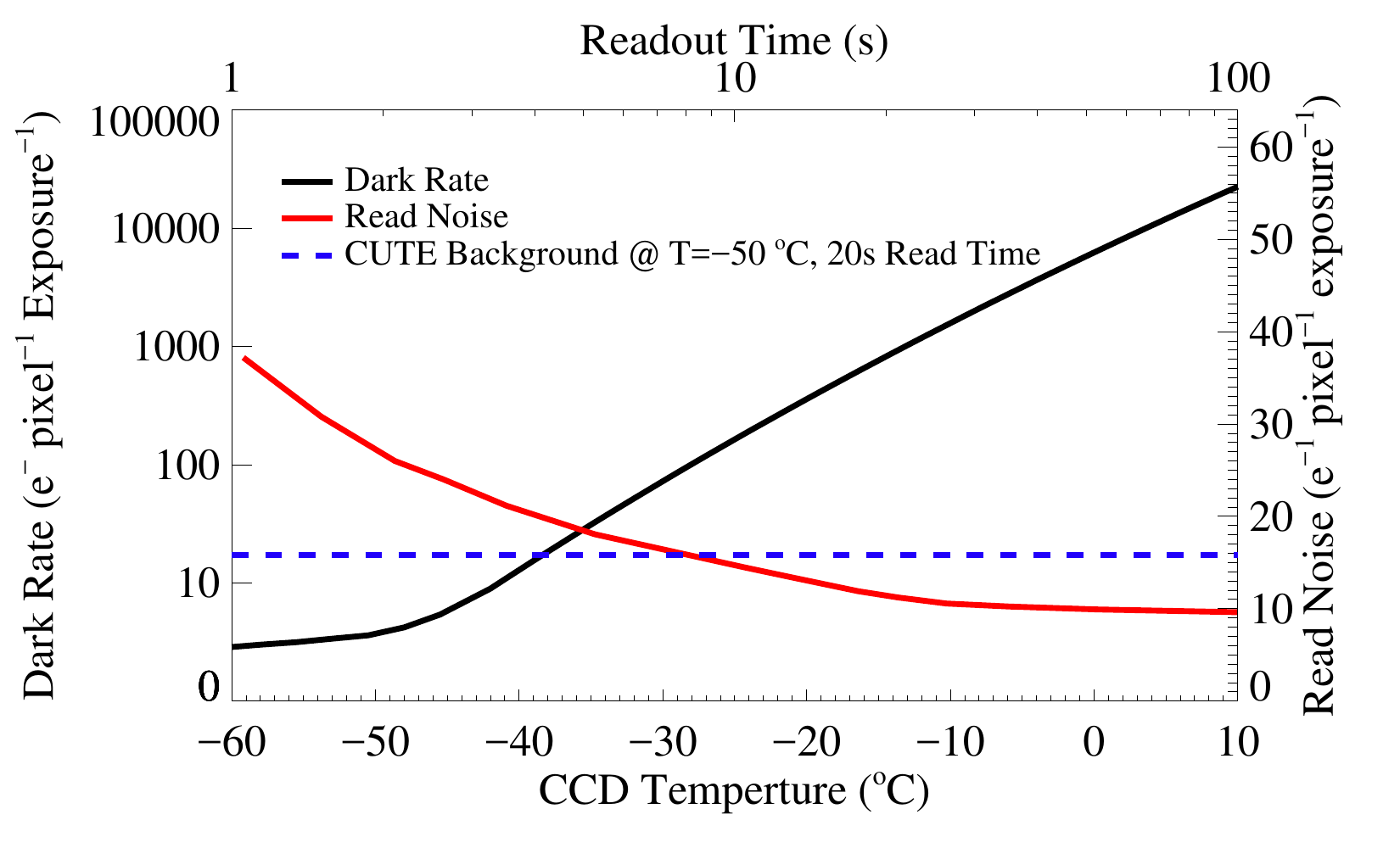}
\end{tabular}
\end{center}
\caption[Trackfits] 
 { \label{fig-CCDback} 
A plot of the CUTE dark per pixel per 300 second exposure as a function of
temperature (black), the counts per
pixel as a function of readout time (red, note: not RMS read
noise, but counts per pixel), and the total CUTE
background per pixel per exposure for -50\adeg\ C and 20 second readout (blue dash, referenced to the read
noise axis).}
\end{figure}

The spectrum is imaged onto an e2v CCD42-10
back-illuminated, UV-enhanced CCD detector. The CCD42-10 has an active
area of 27.6 $\times$ 6.9 mm with 2048 $\times$ 515 pixels, each 13.5 $\mu$m square. The CCD42-10 has flight heritage as the sensor used on the Mars Science Laboratory ChemCham LIBS spectrometer \cite{Wiens12}. The CCD
will be cooled to Peltier temperatures between -50 \adeg C $\geq$ T$_{CCD}$ $\geq$ -60
\adeg C by a Marlow RC3-2.5 thermal electric cooler in thermal contact
with a Cu heat sync and
spacecraft radiator system, resulting in a dark background
rate of 1.2 $\times$ 10$^{-2}$ e$^{-}$ pixel$^{-1}$ s$^{-1}$. The read noise is 3.5
e$^{-1}$ pixel$^{-1}$ RMS at the desired CUTE
readout speed of $\sim$ 20 seconds per frame, resulting in a total
background per pixel per exposure of 15.85 counts
(Fig.~\ref{fig-CCDback}). The CCD is mounted near the sun-facing backside of the CUTE spacecraft
and therefore will require thermal blanketing between the cooled
system and the spacecraft bulkhead. The XB1 and associated CCD
electronics, as well as the batteries, will be located at the far
corner of the spacecraft to be as far from the cooled CCD as
possible (Fig.~\ref{fig-render}). 

The University of Colorado Laboratory for Atmospheric and Space
Physics (CU/LASP) is in the process of installing a dedicated S-band
receiver for cubesat downlink communications. CUTE will be equipped
with a BCT software defined radio (SDR) S-band
transmitter with an anticipated downlink capacity of $\sim$ 1
Mbps, and a Spacequest TRX-U UHF transceiver radio (19.2 Kbs) for
nominal spacecraft operation and as a backup. This yields a
projected daily data capacity of $\sim$ 1.4 Gb assuming two $\sim$ 12
minute passes per day and no resource conflicts.  

Full frame readouts of the CCD will only be transmitted to ground
approximately once per day due to data bandwidth
limitations. These full frames will be used for CCD health checks and
to ensure that the spectral extraction routine is functioning
properly. A typical science exposure will be processed on-board by
first determining the location of the spectrum, and then extracting a
sub-frame consisting of approximately 20$\%$ of
the CCD height ($\sim$ 100 pixels, centered on the spectrum). More than 70$\%$ of the spectral energy of
CUTE is contained within a spectral height of three pixels, with $>$ 99.9$\%$
contained within 7 -- 11 pixels at the points of smallest and largest
spectral flaring, respectively. Extraction over 100 pixels
should therefore be more than sufficient for scientific analysis, and
will reduce the maximum daily data rate to less than 0.7 Gb per day,
including overheads -- a rate that could be covered by a single data
pass. An on-board processing system that extracts the spectrum and
generates a 1D data table will be developed in the event that the
S-band transmitter fails and the slower UHF radio is used. Raw data of the transits will be stored on-board for up to one month and therefore can be
re-analyzed as a full-frame if needed. 

\section{Instrument Performance Estimates}\label{performance}

\begin{figure}[t]
   \begin{center}
   \begin{tabular}{c}
\includegraphics[height=0.3\textwidth,angle=0,trim={0.27in 0.15in 0.35in 0.3in},clip]{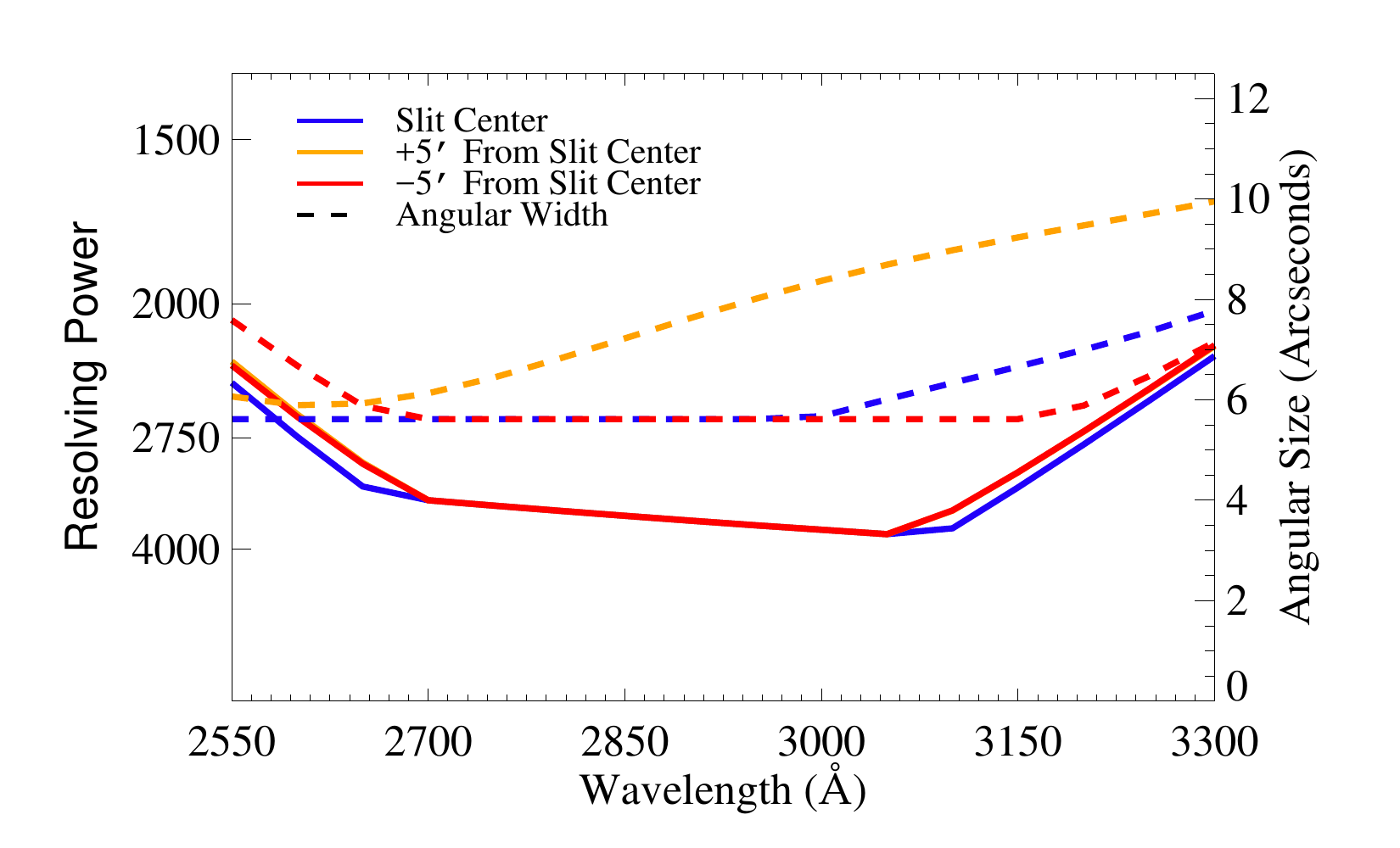}
   \end{tabular}
   \end{center}
   \caption[example] 
  { \label{fig-rpower} 
Spectral resolving power and angular resolution of CUTE as a
function of wavelength for the center and $\pm$ 5\amin\ of the FOV.}
   \end{figure} 

\begin{figure}[b]
   \begin{center}
   \begin{tabular}{c}
   \includegraphics[height=0.3\textwidth,angle=0,trim={0.45in 0.15in 0.75in 0.3in},clip]{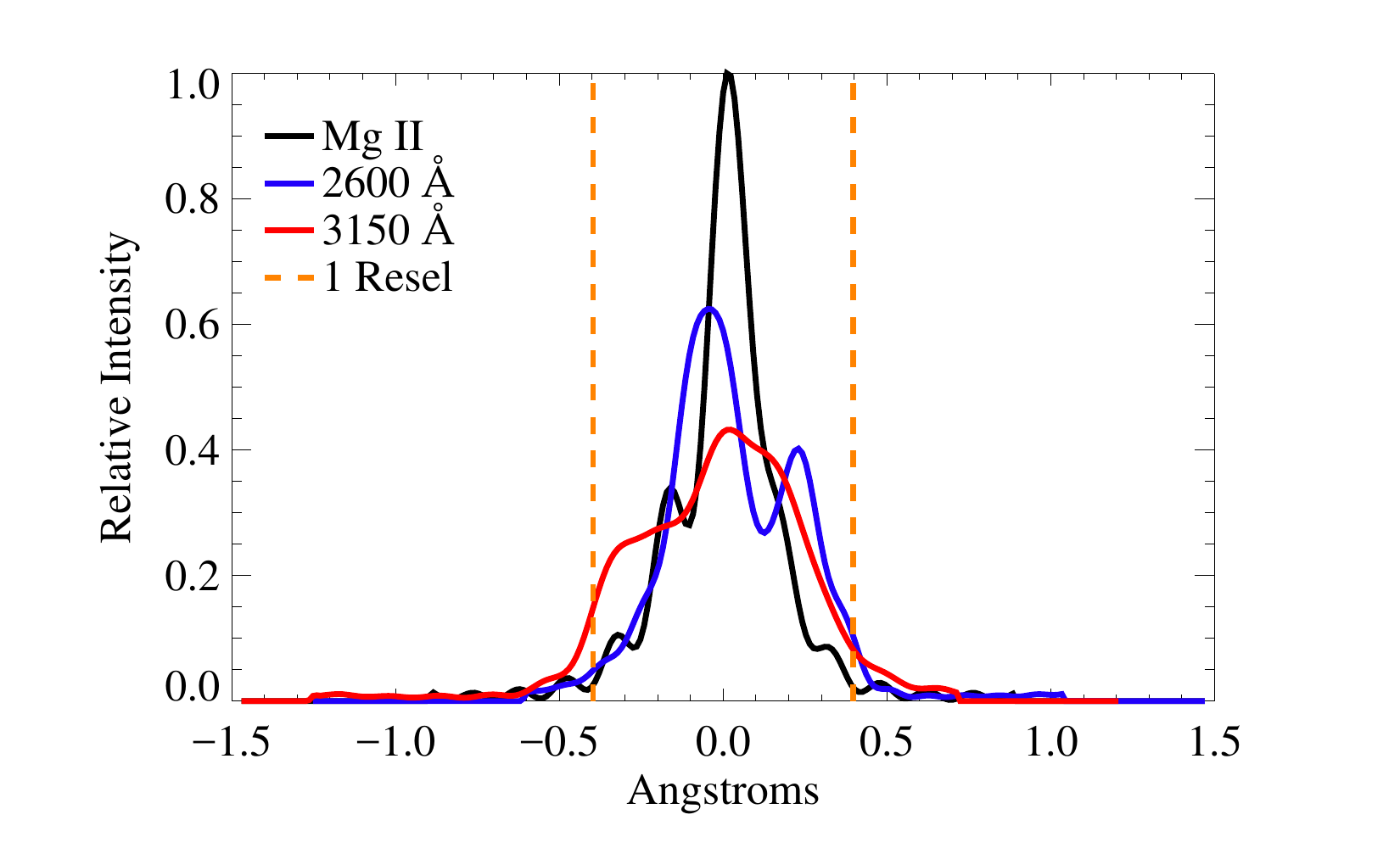} 
   \end{tabular}
   \end{center}
   \caption[example] 
  { \label{fig-lsfs} 
Selected spectral line spread functions (LSFs) of CUTE, with the width
of one resolution element overplotted (dashed lines).}
   \end{figure}

The magnifying imaging spectrograph design follows that of the University of
Colorado sounding rocket payload SISTINE \cite{fleming16}. With a
final focal ratio of F/5.5 and a platescale of 208\asec\ mm$^{-1}$,
CUTE achieves an angular resolution of 5.6\asec\ and an average
resolving power of greater than 3000 across the operational
bandpass. The spectral resolving power is limited by
the size of the CCD resolution elements (defined as two 13.5 $\mu$m
pixels) from 2650 -- 3100 \AA, and the angular resolution is similarly
detector limited from 2550 -- 3000 \AA, meaning that the RMS spot width
at each wavelength in the dispersion and x-dispersion directions is
smaller than the width of two pixels, or one resel (Fig.~\ref{fig-rpower}). Greater
than 70$\%$ of the energy for any single wavelength is contained
within a 2 $\times$ 2 pixel resolution element from 2600 -- 3100 \AA\ (Fig.~\ref{fig-lsfs}). The pointing accuracy of the BCT XB1 spacecraft is listed at
7.2\asec, with the RMS jitter during a five minute exposure being
somewhat smaller, but not yet well defined. This is well matched to the 5.6\asec\
scale of a resel and will contribute to a slight degradation
of the final CUTE performance. CUTE utilizes a long slit despite point
source targets so that the target can be moved around on the
CCD chip in the event of radiation or other damage over the
course of the mission. 

\subsection{Effective Area}\label{aeff}

The effective area of CUTE is the product of the clear collecting area
of the rectangular telescope (A$_{T}$) and the individual component
efficiencies:

\begin{equation}\label{eqn-aeff}
A_{eff}(\lambda) = A_{T} R (\lambda)^{5} \epsilon_{g}(\lambda)
D_{QE}(\lambda)
\end{equation}

\noindent where R($\lambda$) is the reflectivity of the magnesium
fluoride (MgF$_{2}$) protected aluminum
coated optics, $\epsilon_{g}$($\lambda$) is the grating
efficiency, and D$_{QE}$($\lambda$) is the quantum efficiency of the
e2v CCD42-10 detector as reported by e2v. $\epsilon_{g}$($\lambda$) is the modeled blazed grating
efficiency curve for the CUTE grating as predicted by Horiba J-Y. The
peak efficiency of 65$\%$ is comparable to the peak efficiency achieved for similar ion-etched blazed holographic
gratings on $HST$-COS in the FUV (the NUV COS gratings had a unique
ruling pattern that led to reduced efficiency) \cite{Osterman02}. 

\begin{figure}
\begin{center}
\begin{tabular}{c}
\includegraphics[height=0.3\textwidth,angle=0,trim={0.5in 0.15in 0.33in 0.3in},clip]{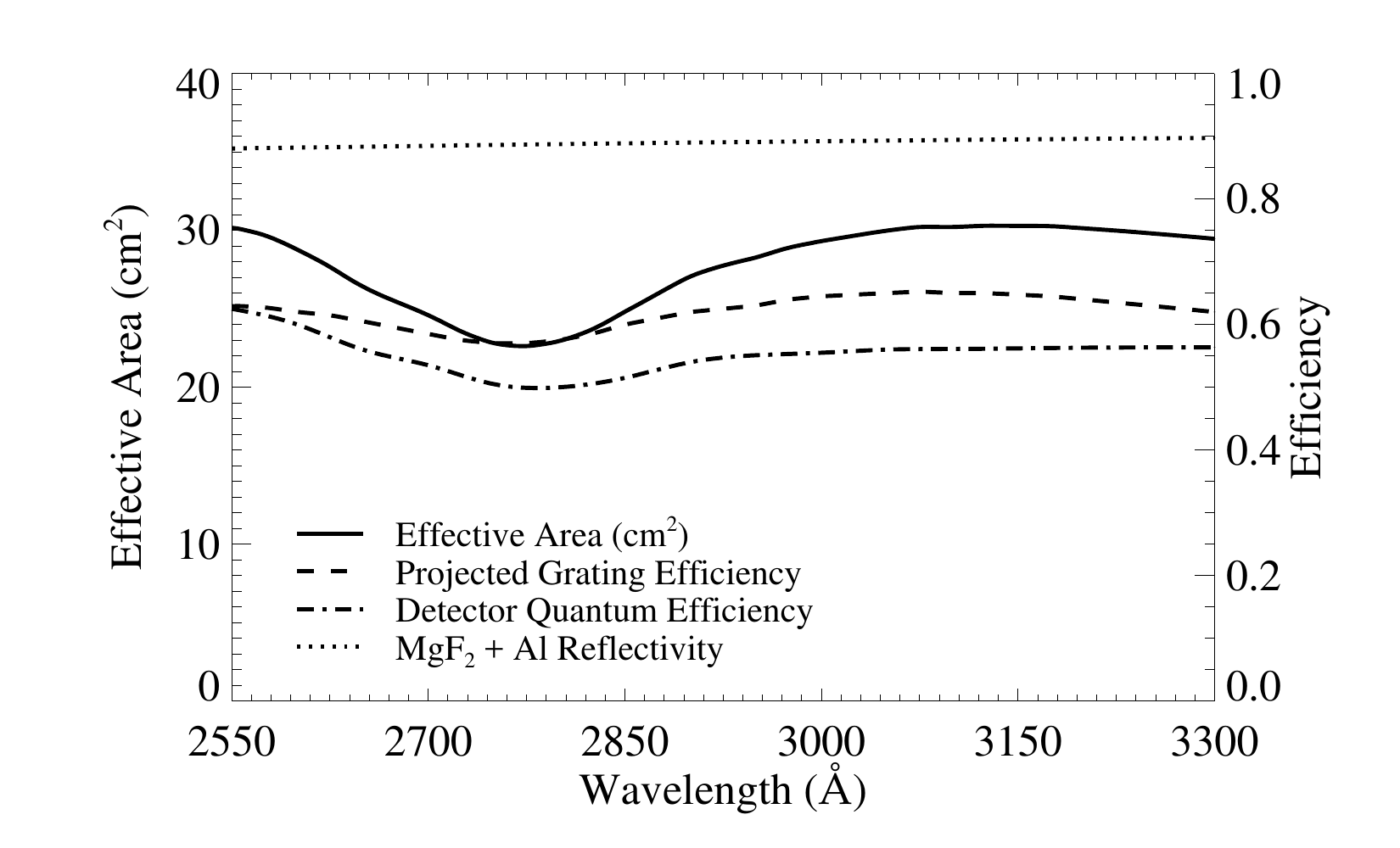}
\end{tabular}
\end{center}
\caption[Trackfits] 
 { \label{fig-aeff} 
The Effective Area curve for CUTE derived from Eqn~\ref{eqn-aeff} with component efficiencies overplotted.}
\end{figure}

CUTE will utilize MgF$_{2}$+Al as opposed to bare aluminum for the optical
coatings in order to prevent the formation of an oxide layer
(Al$_{2}$O$_{3}$) on the aluminum. Oxide formation can interfere with
grating performance, which can result in efficiency
degradation. The CUTE MgF$_{2}$+Al optical coatings
on the grating and second fold mirror will be applied by the NASA
Goddard Space Flight Center (GSFC) Thin Films Coating Laboratory, while Nu-Tek will deliver a coated
telescope and first fold mirror. With properly optimized thicknesses
the GSFC process has been shown to produce coatings with reflectivities
as high as 94$\%$, however we chose to model A$_{eff}$ with a more
conservative R($\lambda$) $<$ 90$\%$. The combined predicted CUTE performance curve with
component efficiencies is presented in Figure~\ref{fig-aeff}.

\subsection{Sources of Systematic Uncertainty}\label{systematics}
Transit observations require a robust understanding of systematic
error, as instrumental or stellar variability over a $\sim$ 2 hour
transit could significantly skew results. The CUTE team, led by
CU/LASP and the Space Research Institute of the
Austrian Academy of Sciences, has begun developing a data simulator to
analyze the impact of potential noise sources (thermal fluctuations, scattered
light, cosmic rays, etc). This simulator will help drive a robust
instrument calibration and testing program at CU/LASP to characterize
the CUTE instrument on the ground and in orbit prior to and during
science operations. 

Many potential systematic errors inherent to transit spectroscopy were
taken into account when the CUTE mission was designed. Certain astrophysical sources, such as
stellar variability, will be mitigated by the sheer observation time
available. CUTE will observe approximately half of a full orbital
phase of each planet target for each transit, providing hours of
baseline for each host star. Random or unexpected short-term events,
such as residual cosmic rays or interloping objects, will likewise be
mitigated by the number of transits observed ($\gtrsim$ 10 per
target), as the data for any outlier transit will be downloaded in
full and studied in more detail. The sun-synchronous orbit of CUTE
will put the spacecraft into a state of near-thermal equilibrium, limiting thermal variability on both the optical bench and
detector. Instrumental performance degradation on-orbit and spacecraft pointing
errors (resulting in the spectrum being imaged on different potions of
the detector) will be accounted for by a robust on-orbit calibration
program. Up to one full day of observation time per week is currently
available for calibration if needed, including observing standard stars at any
detector location. A more detailed analysis will be possible once
the CUTE instrument has been fabricated. 

\section{Science Operations}\label{sci_op}
\begin{figure}[!t]
   \begin{center}
   \begin{tabular}{c}
 \includegraphics[width=0.98\textwidth,angle=0]{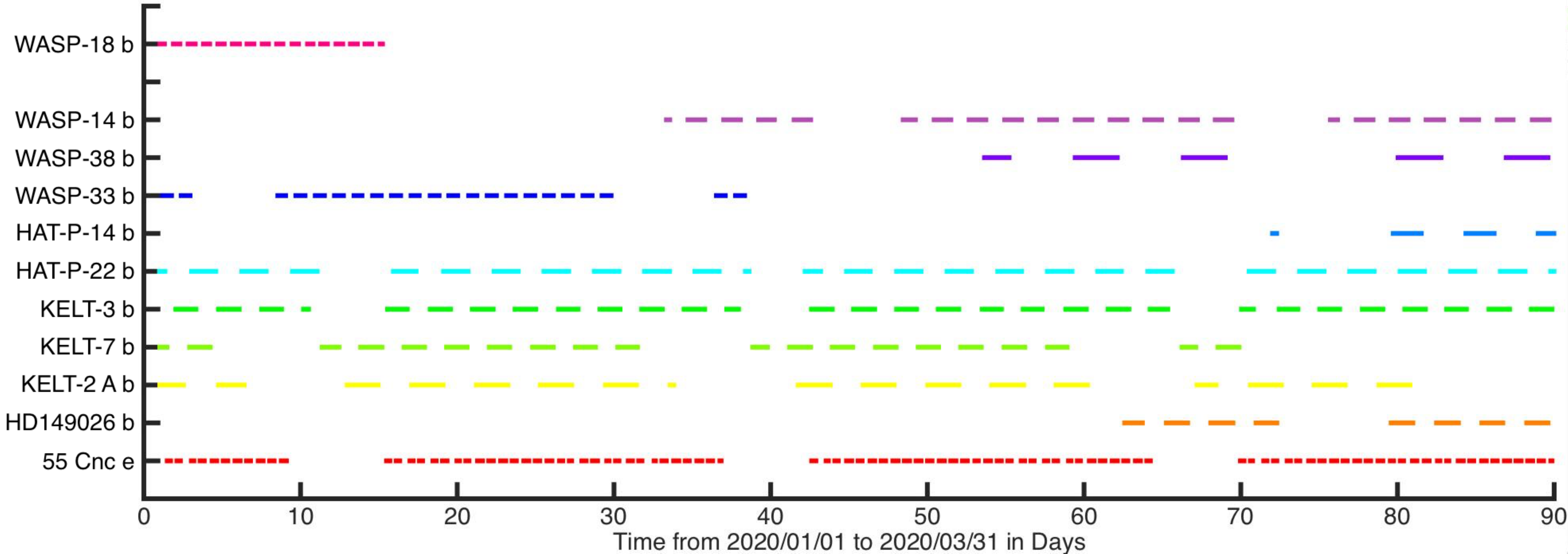}
   \end{tabular}
   \end{center}
   \caption[example] 
  { \label{fig-obswindow} 
Transit visibility windows for a sun-synchronous orbit in a three
month period starting Jan. 1, 2020 for a selection of possible CUTE
targets. The transit phase sampled is -0.25 $<$ $\phi$ $<$ 0.25. Gaps
are due to the planet being out of transit, or the earth, sun or moon violating an avoidance
angle condition. }
   \end{figure}

CUTE will have a nominal operational lifetime of seven months, with
the first month consisting of health checks and on-orbit
calibrations. During the six month primary science mission CUTE will observe
exoplanet transits of hot Jupiters around a sample of $\sim$ 12 bright
stars (out of a sample of up to 30
candidate stars within reach of CUTE) with a broad range of
spectral types (See Table~\ref{tbl-targets} for a subset of possible targets). The list of candidate stars
will increase as new discoveries are announced by ground and space
surveys (e.g. KELT, TESS, MASCARA). Each
target will be prioritized based on the number of available transits during the operations
window, the signal-to-noise of the transit, and the scientific
relevance of the target. It is anticipated that CUTE will observe 10
transits from each of 12 unique targets during the six month
mission lifetime in a sun-synchronous orbit, with a lesser number of transits from other targets
interspersed when there are no higher priority targets available
(Figure~\ref{fig-obswindow}). 

\begin{table}[!b]\scriptsize
\centering
\caption{CUTE Preliminary Representative Target List}
\label{tbl-targets}
\begin{tabular}{lcccccccccccc}
\hline \hline
Target & m$_{V}$ & R$_{P}$ & Period & a$\_$orb & RA & DEC & T$_{eff,*}$ & $\dot{M}_{P}$ & Drag & S/N & S/N  & S/N \\
 &  & (R$_{J})$ & (Day) &(AU) & & & (K) & (kg/s) & (kg/s) & (Mg II) & (Mg I)  & (Cont.) \\
\hline
HD-209458 & 7.7 & 1.36 & 3.53 & 0.047 & 22:03:10.8 & +18:53:03.7 & 6065 & 10$^{8}$ & 9$\times$10$^{6}$ & 102.2 & 72.1 & 204.2 \\
HD-189733 & 7.7 & 1.14 & 2.22 & 0.031 & 20:00:43.7 & +22:42:41.3 & 5040 & 6$\times$10$^{8}$ & 2$\times$10$^{7}$ & 61.4 & 28.5 & 104.8 \\
WASP-33 & 8.3 & 1.50 & 1.22 & 0.026 & 02:26:51.1 & +37:33:01.8 & 7430 & 3$\times$10$^{8}$ & 3$\times$10$^{7}$ & 115.7 & 88.5 & 180.8 \\
KELT-7 & 8.5 & 1.53 & 2.73 & 0.044 & 05:13:10.9 & +33:19:05.8 & 6789 & 10$^{8}$ & 10$^{7}$ & 101.4 & 77.8 & 159.3 \\
WASP-18 & 9.4 & 1.27 & 0.94 & 0.020 & 01:37:25.0 & -45:40:40.5 & 6400 & 2$\times$10$^{7}$ & 10$^{8}$ & 61.2 & 47.8 & 98.7 \\
HAT-P-22 & 9.7 & 1.08 & 3.21 & 0.041 & 10:22:43.6 & +50:07:42.0 & 5302 & 2$\times$10$^{7}$ & 3$\times$10$^{7}$ & 13.5 & 5.6 & 25.4 \\
WASP-74 & 9.7 & 1.56 & 2.14 & 0.037 & 20:18:09.3 & -01:04:32.6 & 5990 & 3$\times$10$^{8}$ & 10$^{7}$ & 26.8 & 19.3 & 62.5 \\
WASP-14 & 9.8 & 1.28 & 2.24 & 0.037 & 14:33:06.4 & +21:53:40.9 & 6475 & 10$^{7}$ & 10$^{8}$ & 48.4 & 38.2 & 79.2 \\
WASP-8 & 9.8 & 1.04 & 8.16 & 0.080 & 23:59:36.1 & -35:01:52.8 & 5600 & 4$\times$10$^{7}$ & 3$\times$10$^{7}$ & 25.1 & 18.1 & 58.9 \\
XO-3 & 9.9 & 1.22 & 3.19 & 0.048 & 04:21:52.7 & +57:49:01.8 & 6429 & 4$\times$10$^{6}$ & 2$\times$10$^{8}$ & 43.4 & 34.4 & 71.5 \\
WASP-69 & 9.9 & 1.06 & 3.87 & 0.045 & 21:00:06.2 & -05:05:40.1 & 4700 & 10$^{9}$ & 3$\times$10$^{6}$ & 12.0 & 5.0 & 22.7 \\
KOI-13 & 9.9 & 1.41 & 1.76 & 0.034 & 19:07:53.1 & +46:52:05.9 & 7650 & 10$^{8}$ & 10$^{7}$ & 42.2 & 33.5 & 69.7 \\
55-Cnc & 6.0 & --- & 14.65 & 0.11 & 08:52:36.1 & +28:19:53.0 & 5196 & 3$\times$10$^{7}$ & 10$^{7}$ & 157.7 & 78.7 & 255.3 \\
\hline
\end{tabular}\\
\raggedright
\vspace{-0.1in}
\end{table}
\normalsize

While the nominal mission is only seven months, the CUTE satellite will have an orbital
lifetime of at least one year and could be extended until orbital
decay. The nominal mission will be extended automatically after
deployment until at least the end of the funding period of performance
(June 2021), as much as
1.5 years if CUTE were to launch January 2020 as planned. Therefore, while we baseline at least 100 transits in the
nominal mission, this number could more than double depending on the
launch date and longevity of the spacecraft.

CUTE will observe each transit across half the exoplanet orbital phase (-0.25
$<$ $\phi$ $<$ 0.25) to ensure a well
established stellar flux baseline. It is not certain when the
early ingress transit will begin, as there is no foreknowledge of the
extent of the atmospheres for most targets, therefore the wide phase
range will ensure that the ingress/egress will be captured. With an
average planetary orbital period of three days, CUTE will typically
observe 3-5 transits per week. Each primary target will be visited for multiple transits to ensure complete transit
coverage and to search for variability in the transit shape and
depth. Cute will have greater than a 70$\%$
observing efficiency for most targets in the anticipated
sun-synchronous orbit, which will enable the majority of the light
curve for each targeted transit to be observed continuously. CUTE is
also compatible with an ISS or other equatorial orbit, however Earth
occultation will reduce observation efficiency to $\lesssim$ 40$\%$ and
require multiple transit observations to construct a full light
curve. We anticipate only five transits of 12 targets as a baseline
mission should CUTE be launched into such an orbit, as well as fewer
ground station contacts per day, as an equatorial orbit will not pass
over the second LASP ground station in Fairbanks, Alaska. A proposal to the
NASA Cubesat Launch Initiative has been submitted requesting that CUTE
be placed on the cubesat launch manifest with a
sun-synchronous orbit.

Students at the University of Colorado are currently developing an
algorithm to optimize the operational efficiency of CUTE. This algorithm projects the
availability of all targets on the sky for a given orbit, the
transit phase, and the scientific relevance to select the most
valuable target for observation at any given time
(Figure~\ref{fig-obswindow}). Exclusion angles for the sun, moon and
Earth are all accounted for. The science team
will then select from the ranked targets an observing plan on a
week-by-week basis, updated as data is analyzed and the science
priorities are reassessed. We find that CUTE will be capable of
meeting the goal of at least 10 transits of 12
systems within the baseline mission lifetime. 

\subsection{Scientific Capabilities}\label{sci_cap}
\begin{figure}[t]
   \begin{center}
   \begin{tabular}{c}
   \includegraphics[height=0.35\textwidth,angle=0,trim={0.00in 0.00in 0.0in 0.0in},clip]{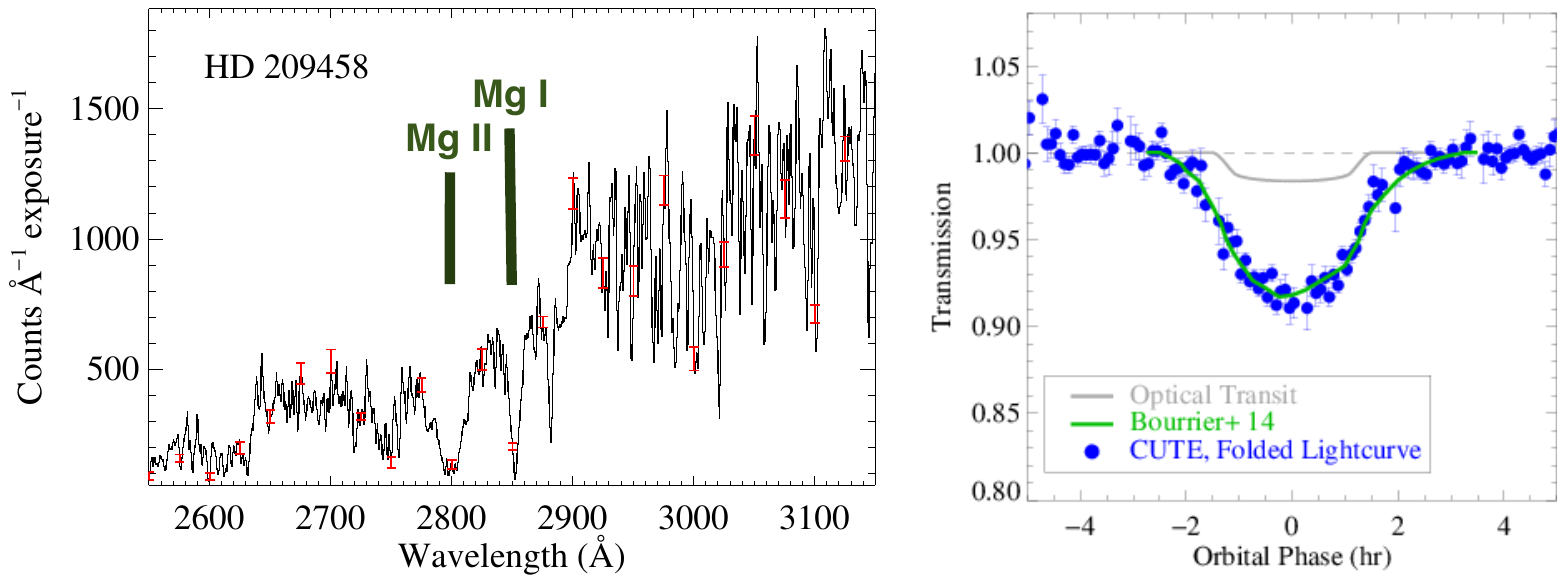} 
   \end{tabular}
   \end{center}
   \caption[example] 
  { \label{fig-countsim} 
(Left) Simulated 300 second CUTE spectrum of HD 209458.
(Right) The folded lightcurve of 10 simulated \ion{Mg}{1} transits of HD 209458b as
observed by CUTE, based on the model of Bourrier et al. 2014
\cite{Bourrier14}. The background and simulated spectrum from the
left is folded into this simulation.}
   \end{figure}

Figure~\ref{fig-countsim} shows a simulated 300 second CUTE spectrum of HD
209458, one of the brighter stars in the CUTE sample (V$_{mag}$ = 7.6,
T$_{eff}$ = 6000 K, Table~\ref{tbl-targets}). These count rates are
consistent with an average SNR $>$ 18 per resolution element based on the backgrounds projected in
Figure~\ref{fig-CCDback}. The integrated absorption regions near \ion{Mg}{2} and \ion{Mg}{1} will likewise have a SNR $>$ 50, making
CUTE sensitive to transit depths as low as 0.1 -- 1$\%$ when folded over
two or more transits. With sufficient signal-to-noise folded over
multiple transits it may be possible to resolve other individual
atmospheric absorptions lines/bands. The transit
depth is expected to be 4~--~10$\%$ in atmospheric tracers such
as around \ion{Mg}{1} and \ion{Mg}{2} \cite{Fossati10, Haswell12, Bourrier14}. 

We simulated CUTE observations of ten transits of HD 209458b with an
ISS orbit (40$\%$ covering fraction per orbit), assuming the 
\ion{Mg}{1} transit model of Bourrier et al. (2014).  CUTE samples
the entire light curve from -0.25 $<$ $\phi$ $<$ 0.25 over the ten
transits with the precision to sample the ingress and egress
of an extended atmosphere (Fig.~\ref{fig-countsim}). The projected SNR
for a band of spectrum within $\pm$ 10 \AA\ of the center of the neutral and singly
ionized Mg lines, as well as for a 50 \AA\ band of the continuum,
for a subset of possible CUTE targets is presented in
Table~\ref{tbl-targets}. We integrate over a large bandpass rather
than fit the \ion{Mg}{2} line profiles to encompass multiple singly ionized gas
species around the \ion{Mg}{2} feature
\cite{Fossati10,Nichols15}. Similarly, we integrate over a large
``line-free'' continuum region near 2900 \AA\ as a comparison point
with optical transits.  For specific bright, early-type targets (e.g., KELT-9b), 
high-SNR ($>$~50 per spectral resolution element) spectra can be acquired
in a single transit observation.

\subsection{Science Closure}
CUTE is anticipated to provide $\sim$ 10 transit light curves ($>$
60$\%$ coverage over a -0.25 $\leq$ $\phi$ $\leq$ 0.25 phase range) of
approximately twelve short-period exoplanets by the end of the nominal
seven-month mission. A number of other planetary systems will be observed per
the discretion of the science team during scheduling gaps (see
Fig.~\ref{fig-obswindow}). This data will represent more than an order
of magnitude increase in the number of spectroscopic NUV transits
observed to-date, bringing the observational basis for atmospheric
escape studies on par with hydrodynamic models of exoplanet
atmospheric mass loss (\S\ref{sciback}) \cite{Koskinen07, Koskinen13a,
  Koskinen13b, Bourrier13}. CUTE will help
to resolve the ingress variability observed in systems
like WASP-12b by covering a large number of transits per
system \cite{Haswell12,Nichols15}, and likewise address the effect of
stellar variability in the data by significant out-of-transit monitoring. Measurements of this breadth are not
feasible with a shared resource like $HST$, the only existing
observatory capable of these observations. Complimentary ground-based spectropolarimetric observations of
the stellar magnetic field could potentially provide information on
the planetary magnetic field as well \cite{Vidotto11a}. The expected on-orbit lifetime
of CUTE is 2 years, enabling a possible extended mission, increasing
the final CUTE database to 24 -- 30 close-in planets. With
all data products publicly available, we expect CUTE to make a
significant contribution to the study of atmospheric mass loss in
exoplanet systems. 

\section{CUTE Development and Testing Schedule}\label{schedule}

\begin{figure}[!t]
   \begin{center}
   \begin{tabular}{c}
   \includegraphics[width=1.0\textwidth,angle=0,trim={0.0in 1.35in 0.2in 3.5in},clip]{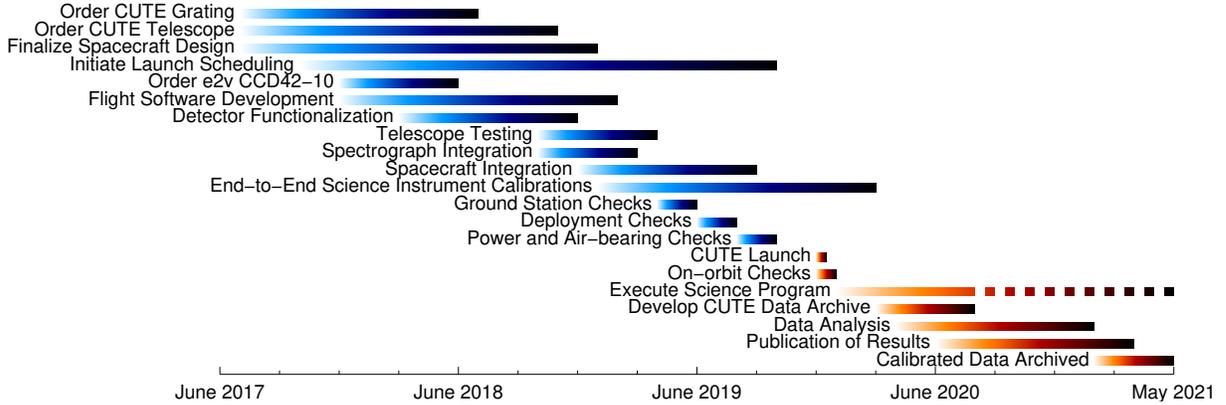}
   \end{tabular}
   \end{center}
   \caption[example] 
  { \label{fig-schedule} 
Anticipated CUTE fabrication and launch schedule. }
   \end{figure}

Funding for the fabrication of CUTE began July 2017 and orders
are being assembled for long lead time items, including the
grating and telescope optics (Fig.~\ref{fig-schedule}). The grating
has an estimated delivery date from Horiba J-Y of March 2018, while
we anticipate that the assembled, aligned and focused telescope will
be delivered by Nu-Tek in late 2018. Following delivery of the telescope, we
will perform a quality review that will include throughput testing,
spot size measurement, and another vibration test. The CCD42-10, as
well as two engineering chips, will
be delivered in
early 2018 and the process of creating a functional controller and
mount for the CCD
chip and designing and testing the thermal mount will proceed
throughout 2018. The grating will be delivered coated in platinum for preliminary efficiency and
dispersion measurements in the CU square tank facility, after which it
will be sent to GSFC for a MgF$_{2}$+Al coating. After a post-coating reflectivity measurement,
the spectrograph will be installed onto the telescope (see
\S\ref{telescope}). Spectrograph focusing and alignment will be
achieved via a piston-tip/tilt grating mount. The assembled science
instrument will be illuminated with collimated light from a D$_{2}$ or
PtNe lamp installed onto the
CU Long Tank facility for final system alignment using the science detector. 

There will be a direct path to the first fold mirror for a region of
sky subtending a total of 1310 square degrees due to the large aperture and fast primary
mirror (F/0.75). While not all of this will reflect into the slit, we
assume all of it does for the sake of estimating the ``worst-case'' scattered light
background. We assume a conservatively high flux density of
10$^{-7}$ erg cm$^{-2}$ \AA$^{-1}$ s$^{-1}$ deg$^{-2}$ in the NUV, or roughly the
equivalent of one Sirius deg$^{-2}$. The CUTE spectrograph will be
fabricated with light trapping baffles machined into the mount
structure that will require a minimum of three bounces for a photon
incident on the baffle to escape on a vector towards the next optical
element, and at least two such reflections for off-axis light to align
sufficiently with the optical path to reach the detector, which faces
away from the slit (Fig.~\ref{fig-trace}). The light traps will be
fabricated from a blackened material,
such as roughened black delrin, invar or black anodized and roughened
aluminum with $<$ 3$\%$ reflectivity in the near UV, suppressing the scattered
light by a $minimum$ of nine orders of magnitude (0.03$^{6}$ $<$ 10$^{-9}$). This
will limit the photon flux to $<$ 50$\%$ of the dimmest CUTE target,
which when isotropically distributed around the detector will
represent a background of $<$ 1$\%$ per resel for this worst-case scenario. Rigorous
scattered light testing will be carried out at CU using a Hg
``pen-ray'' to illuminate the telescope aperture with intense off-axis NUV
radiation.   

End-to-end testing will proceed starting in mid 2019 after integration
into the XB1 spacecraft. CUTE will be
subjected to thermal vacuum testing in CU/LASP facilities to
demonstrate system survivability and optical stability. Handshaking and
communications testing will proceed in late 2019 via the LASP ground
station, which has successfully served as the Mission Operations
Center (MOC) for other LASP-built
cubesats utilizing the XB1 system, including MinXSS
\cite{Mason17}. The target launch date for CUTE will be in early 2020 to any orbit
visible from the CU ground station, including an ISS or a
sun-synchronous orbit. Nominal science operations will be carried out
for at least seven months from deployment, with one month of on-orbit
commissioning and six months of science operations. A data reduction pipeline
is being designed by the Space Research Institute of the Austrian
Academy of Sciences to handle the on-board data processing, including
the background subtraction, cosmic-ray rejection and flat fielding of each CCD frame and the
generation of a 1-D spectrum for transmission to the ground. A portion
of the raw data will also be transmitted and analyzed on the ground (\S\ref{detector}). Calibrated, background subtracted one-dimensional spectral data
will be made publicly available on CU-LASP servers at the conclusion of
the mission. 

\section{Summary}
The University of Colorado UV instrumentation group at the Laboratory
for Atmospheric and Space Physics has finalized the design of the Colorado
Ultraviolet Transit Experiment and will begin fabrication in 2018. CUTE is a low resolution (R $\approx$ 3000) NUV imaging
spectrograph with spectral coverage of important exoplanet atmospheric
tracers of Mg, Fe, and OH. A rectangular aperture results in
greater than three times the collecting area relative to a circular
aperture in a 6U cubesat form factor. The average effective area of CUTE
is 28 cm$^{2}$, or $\approx$ 70$\%$ of the effective area (and $\approx$ 40$\times$
the resolving power) of the GALEX NUV
grism, delivered in a 6U cubesat package. The high quality spectral
and angular resolution, which produce monochromatic point source
images smaller than a single resolution element on the detector, are
made possible by a multi-passed optical design and a blazed, ion-etched
aberration correcting holographic grating. CUTE will be the first NASA funded UV/O/IR cubesat for scientific
astronomy, and is designed to demonstrate that high quality science data
can be obtained from an instrument in a cubesat package and a
sub-orbital class budget. The target
launch date for CUTE is early 2020, after which CUTE will embark on a
nominal seven month mission to monitor the transits of hot Jupiter
exoplanets to quantify atmospheric mass loss and magnetic fields.

\acknowledgments  
The authors would like to thank Nu-Tek Precision Optics,
Horiba Jobin-Yvon, and Blue Canyon Technologies for their helpful discussions during the
design phase. This work is funded by NASA grant number NNX17AI84G to the University
of Colorado. 

{\bf Brian Fleming} is an assistant research professor at the University of
Colorado, Boulder and NASA Nancy Grace Roman Technology Fellow specializing in space instrumentation and energetic radiation escape from galaxies. He is the optical designer and project scientist for a
number of NASA sub-orbital programs and serves on the instrument
definition team for the LUVOIR UV spectrograph LUMOS.

{\bf Dr. Kevin France's} research focuses on exoplanets and their host
stars, protoplanetary disks, and the development of instrumentation
for ultraviolet astrophysics. He was a member of the HST-COS
instrument and science teams, a voting member of the LUVOIR STDT, is a
regular guest observer with the Hubble Space Telescope, and is the PI
of NASA-supported sounding rocket and small satellite programs to
study exoplanet atmospheres and flight-test critical path hardware for
future UV/optical missions.   

{\bf Luca Fossati} is a junior group leader at the Space Research Institute of the Austrian Academy of Sciences studying exoplanet atmospheric escape.

{\bf Tommi Koskinen} is an assistant professor at the Lunar and Planetary Laboratory of the University of Arizona.  He is an expert on the physics and chemistry of the middle and upper atmosphere.  His research combines the analysis of remote sensing data with numerical modelling to characterize planetary atmospheres in a wide range of different environments, ranging from the solar system to distant exoplanetary systems.

{\bf Jean-Michel Désert} is an Assistant Professor of Astronomy and Astrophysics at the University of Amsterdam (UvA) in the Netherlands studying exoplanets and their atmospheres.

{\bf Arika Egan} is an Astrophysics and Planetary Science graduate student at the University of
Colorado Boulder working on the CUTE cubesat project. 

{\bf Nicholas Nell} is an Aerospace Engineering graduate student at
the University of Colorado working on the SISTINE sounding rocket
project. He formerly served as the electrical engineer for the CU UV
astronomy instrumentation group. 

{\bf Kelsey Pool} is an undergraduate student at the University of
Colorado Boulder studying Mechanical Engineering. She is currently a
research assistant on the CUTE cubesat project. 

No other biographies are available. 


\bibliography{./mybib} 
\end{spacing}

\end{document}